# Spectral nature of soiling and its impact on multi-junction based concentrator systems

Eduardo F. Fernández[1*], Daniel Chemisana[2], Leonardo Micheli[1] and Florencia Almonacid[1]

[1] *Centre for Advanced Studies on Energy and Environment (CEAEMA), University of Jaén, Jaén 23071, Spain*

[2] *Applied Physics Section of the Environmental Science Department, University of Lleida, Lleida 25001, Spain*

*Corresponding author: eduardo.fernandez@ujaen.es, Tel: + 34 953 21 3520*

**Abstract:**
Soiling, which consists of dust, dirt and particles accumulated on the surface of conventional or concentrator photovoltaic modules, absorbs, scatters, and reflects part of the incoming sunlight. Therefore, it reduces the amount of energy converted by the semiconductor solar cells. This work focuses on the effect of soiling on the spectral performance of multi-junction (MJ) cells, widely used in concentrator photovoltaic (CPV) applications. Novel indexes, useful to quantify the spectral impact of soiling are introduced, and their meanings are discussed. The results of a one-year experimental investigation conducted in Spain are presented and are used to discuss how soiling impacts each of the subcells of a MJ cell, as well as the cell current-matching. Results show that soiling affects the current balance among the junctions, i.e. the transmittance losses have found to be around 4% higher in the top than in the middle subcell. The spectral nature of soiling has demonstrated to increase the annual spectral losses of around 2%. Ideal conditions for the mitigation of soiling are also discussed and found to be in blue-rich environments, where the higher light intensity at the shorter wavelengths can limit the impact of soiling on the overall production of the CPV system.

## 1. Introduction

The accumulation of dust, dirt and particles, commonly referred to as "soiling", is one of the most relevant concerns regarding the performance of photovoltaic (PV) systems. Soiling reduces the irradiance that reaches the semiconductor materials responsible of the conversion of the sunlight into electricity. In this sense, soiling effects can cause power losses up to 70% in the worst scenarios [1]. Bearing this in mind, several studies have investigated the causes of soiling and their relationships with the environmental parameters [2, 3, 4, 5]. Other studies have also pointed out that the attenuation and the scattering phenomena produced by soiling present a clear wavelength dependence [6, 7]. Indeed, it has been found that soiling produces a higher attenuation in the blue region, and therefore, causes a red-shift of the incident spectral distribution. Consequently, the impact of soiling is going to depend on the spectral response of the specific PV material under investigation [8, 9]. In any case, this type of studies are scarce and further efforts are still needed to fully understand the relation between the spectral transmittance of soiling and the absorption band of PV devices.

Concentrator PhotoVoltaics (CPV) has achieved the highest conversion efficiencies among all the PV technologies, and has a noteworthy potential to deliver high-energy yields and low-cost electricity at locations with high solar energy resource [10]. The use of III-V multi-junction (MJ) solar cells maximizes the absorption of the incident terrestrial solar spectrum. In addition, the use of concentrator optics reduces the amount of semiconductor material and contributes to increase the overall efficiency of the system [11]. One of the most relevant differences between the performance of MJ-based CPV and single-junction PV systems is related with their spectral dependence [12]. The series connection of various semiconductors with different energy gaps makes CPV technology much more sensitive to the key parameters that affect the input spectrum, namely: air mass, aerosols and precipitable water [13, 14]. In this sense, it has been found that MJ-based CPV systems present annual spectral losses around 5%-10% higher than conventional PVs without considering the additional impact due to soiling [15]. Indeed, several doubts raise when trying to develop MJ solar cells with more than four band-gaps due to subcell current limiting issues produced by the inherent spectral variations in outdoors [14].

Bearing the above in mind, it can be expected that soiling modifies the spectral performance of MJ-based CPV systems and introduces additional non-negligible losses in the system. However, to the present, the only study concerning this issue has been conducted in [16]. The authors investigated the impact of soiling by considering artificial soiling and a set of simulated spectra. This study is of great value to understand the spectral effects of soiling and demonstrated for the first time that soiling can affect the current balance among the subcells in a MJ solar cell. However, further investigations are needed to better understand the spectral impact of soiling under real working conditions, where the type and amount of soiling, as well as the characteristics of the spectral irradiance, vary with time.

The present work is intended to fill the gap concerning the spectral nature of soiling and its impact on MJ-based CPV systems as a function of the time-varying input spectrum. To address this issue, the spectral transmittance of soiling and the spectral irradiance recorded over a course of a year in a location in Southern Spain have been used. Based on these data, the effects of soiling on the key wavebands of a typical MJ solar cell are analyzed and discussed. In addition, by using a set of novel equations, the impact of the spectral transmittance of soiling in the current balance of the subcells and in the performance of the system is quantified and discussed in detail for the first time. This study offers the first investigation concerning the spectral impact of soiling in MJ-based CPV systems under real operating conditions. The final goal is to contribute to the understanding of the outdoor performance of CPV systems in the presence of soiling and to find new solutions to mitigate its effect.

**2. Theoretical background**

The impact of soiling on the performance of a PV device can be calculated by means of the Soiling Ratio (SRatio), which can be expressed as [17]:

$$SRatio = \frac{J_{sc,soiled}}{J_{sc,cleaned}} \quad (1)$$

where $J_{sc,soiled}$ and $J_{sc,cleaned}$ are, respectively, the short-circuit current densities of a soiled and a reference cleaned PV device of the same technology exposed under the same conditions. Even if the ratio of the short-circuit currents neglects the influence of non-uniform soiling, it is widely accepted as a suitable approach to quantify the impact of soiling [4]. In the case of MJ-based CPV systems, the current densities of equation (1) can be expressed as [18]:

$$J_{sc,cleaned} = \min(\int_{\lambda_{min}^i}^{\lambda_{max}^i} E_b(\lambda) SR_i(\lambda) d\lambda) = \min(J_{sc,cleaned}^i) \quad (2)$$

$$J_{sc,soiled} = \min(\int_{\lambda_{min}^i}^{\lambda_{max}^i} E_b(\lambda) \tau_{soiling}(\lambda) SR_i(\lambda) d\lambda) = \min(J_{sc,soiled}^i) \quad (3)$$

where $SR_i(\lambda)$ and $J_{sc}^i$ are, respectively, the spectral response and the short-circuit current density of the $i^{th}$-junction, $E_b(\lambda)$ is the spectral distribution of the direct normal irradiance (DNI), $\tau_{soiling}(\lambda)$ is the direct spectral transmittance of the soiling accumulated on the surface of the CPV device, and $\lambda_{max}$ and $\lambda_{min}$ are, respectively, the longest and shortest wavelengths of the $i^{th}$-junction.

The SRatio, as defined in equations (1), (2) and (3), expresses the effects of both the broadband attenuation and the spectral variation of the irradiance profile produced by soiling. The broadband attenuation considers the average reduction of the light intensity in the absorption band of the MJ cell, while the spectral effects are due to the non-flat spectral transmittance profile of soiling as a function of wavelength. If only the broadband effects want to be considered, equation (1) needs to be rewritten as:

$$BSRatio = \frac{\int_{\lambda_{min}^{MJ}}^{\lambda_{max}^{MJ}} E_b(\lambda) \tau_{soiling}(\lambda) d\lambda}{\int_{\lambda_{min}^{MJ}}^{\lambda_{max}^{MJ}} E_b(\lambda) d\lambda} \quad (4)$$

where BSRatio is defined as the Broadband Soiling Ratio, and $\lambda_{max}^{MJ}$ and $\lambda_{min}^{MJ}$ are, respectively, the longest and shortest wavelength of the MJ solar cell. This expression is obtained by combining equations (1) to (3) and considering a flat and perfect spectral response for all the junctions of the MJ cell, i.e. $SR_i(\lambda) = 1$. On the other hand, if only the spectral effects of soiling aim to be considered, equation (1) needs to be rearranged as:

$$SSRatio = \frac{SRatio}{BSRatio} = \frac{\min(\int_{\lambda_{min}^i}^{\lambda_{max}^i} E_b(\lambda) \tau_{soiling}(\lambda) SR_i(\lambda) d\lambda)}{\min(\int_{\lambda_{min}^i}^{\lambda_{max}^i} E_b(\lambda) SR_i(\lambda) d\lambda)} \cdot \frac{\int_{\lambda_{min}^{MJ}}^{\lambda_{max}^{MJ}} E_b(\lambda) d\lambda}{\int_{\lambda_{min}^{MJ}}^{\lambda_{max}^{MJ}} E_b(\lambda) \tau_{soiling}(\lambda) d\lambda} \quad (5)$$

where SSRatio is defined as the Spectral Soiling Ratio. This expression accounts only for the impact of the soiling spectral profile on the performance of MJ-based CPV systems. In this sense, if the current decreases in a larger extent than the irradiance, the SSRatio presents a value lower than one, which means a worse spectral performance. On the other hand, if the current decreases in a lesser extent than the irradiance, the SSRatio presents a value higher than one, which means a better spectral performance. In other words, the SSRatio is higher/lower than one if the attenuation of soiling is lower/higher in those portions of the spectrum where the irradiance and the SR of the subcells are less/more intense.

As mentioned, soiling has been found to produce a higher attenuation at shorter wavelengths. Hence, it is expected to modify the current balance among the subcells of MJ solar cells, and therefore, to affect the spectral performance of the system. The relative spectral impact between the top and middle junctions of a reference cleaned MJ solar cell can be evaluated through Spectral Matching Ratio (SMR) as [19]:

$$SMR_{cleaned} = \frac{J_{sc,cleaned}^{top}}{J_{sc,cleaned}^{mid}} \cdot \frac{J_{sc,cleaned}^{mid*}}{J_{sc,cleaned}^{top*}} \qquad (6)$$

where "*" refers to the current densities under the reference spectrum AM1.5d ASTM G-173-03 at which MJ solar cells and CPV systems are rated [20]. For a soiled device, equation (6) can be rewritten as:

$$SMR_{soiled} = \frac{J_{sc,soiled}^{top}}{J_{sc,soiled}^{mid}} \cdot \frac{J_{sc,cleaned}^{mid*}}{J_{sc,cleaned}^{top*}} \qquad (7)$$

This equation quantifies the possible effects of soiling on the current balance, however, it also considers the inherent effects of the input spectral irradiance. Hence, this expression is not valid if only the spectral effects produced by soiling aim to be investigated. By combining equations (6) and (7), it is possible to define a new metrics denoted as Soiling Mismatch ratio (SMratio) as:

$$SMratio = \frac{SMR_{soiled}}{SMR_{cleaned}} = \frac{J_{sc,soiled}^{top}}{J_{sc,soiled}^{mid}} \cdot \frac{J_{sc,cleaned}^{mid}}{J_{sc,cleaned}^{top}} \qquad (9)$$

The SMratio is a normalized index that accounts for the additional spectral effects produced by soiling on the current balance among the top and middle subcell. In this sense, an SMratio lower than one represents more soiling spectral losses for the top junction respect to the middle junction, (i.e. more losses in the blue region), and a value higher than one represents more soiling spectral losses for the middle junction respect to the top junction (i.e. more losses in the red region).

It worth to mention that the SMratio, as well as the SMR$_{cleaned}$ and SMR$_{soiled}$, could be formulated to evaluate the relative spectral impact among all the junctions of MJ cells made up of three or more junctions. However, CPV systems have being historically be widely based on III-V triple-junction solar cells grown on Germanium substrates [11]. As a consequence, the impact of soiling on the bottom subcell can be discarded because this subcell produces around 30% more current than the top and middle subcells [21]. Hence, it is unlikely for it to become the limiting subcell of the stack because of soiling, given also the limited attenuation that soiling causes at higher wavelengths previously found [6].

Finally, the Average Spectral Transmittance (AST) of soiling across a specific spectral region can be calculated by means of the following relationship [22]:

$$AST_j(\lambda) = \frac{1}{\lambda_{max}^j - \lambda_{min}^j} \int_{\lambda_{min}^j}^{\lambda_{max}^j} \tau_{soiling}(\lambda) d\lambda \qquad (10)$$

where $\lambda_{max}^j$ and $\lambda_{min}^j$ are, respectively, the longest and shortest wavelengths of the $j$<sup>th</sup>-waveband considered. In this work, the spectral regions of a typical lattice-matched GaInP/GaInAs/Ge solar cell have been used. This index accounts for the reduction on the light intensity as a function of wavelength across the desired spectral range. The main difference with the BSRatio is that the later also accounts for the actual spectral distribution of the sunlight.

## 3. Experimental procedure

To conduct this study, a one-year (31/12/2016 to 16/01/2018) outdoor experimental campaign was conducted at the Centro de Estudios Avanzados en Energía y Medio Ambiente (CEAEMA) of the University of Jaén, in Jaén, Southern Spain (latitude 37º49'N, longitude 3º48'W, elev. 457 m). Jaén is a non-industrialized medium-size city with a high annual insolation resource, >1,800 kWh/m$^2$, and wide range of air temperatures that usually go from less than 5ºC in winter to more than 40ºC in summer [23]. It also presents low-medium values of precipitable water and aerosols, although the second ones can periodically reach unusually high values due to specific and stochastic events, such as Saharan dust storms or the burning of olive trees branches in the region. Indeed, Jaén is the largest producer of Olive oil in Spain and worldwide with a 50% and 20% of the total production respectively [24]. Therefore, the present work involves different amounts and types of dust. This increases the interest of the study and opens the way to extract conclusions valid for other locations worldwide. This is in agreement with a recent study of the authors aimed to correlate the total soiling losses of PV technology to different waveband and single-value soiling transmittance measurements [22]. As shown in that work, the main conclusions found in Jaén were found to be also valid in other two locations in the USA, Golden (Colorado) and San José (California).

The procedure to estimate the spectral transmittance of soiling is analogous to the work presented in [22, 6]. One Diamant® low-iron glass coupon 4 cm × 4 cm in size and 3 mm thick from Saint-Gobain Glass was placed horizontally outdoors to capture as much as possible natural dust. This coupon, denoted as *Soiled Coupon*, was never cleaned and its direct transmittance was measured weekly within a wavelength range between 300 and 2,000 nm, using a Lambda 950 UV/Vis spectrophotometer at the Center of Scientific-Technical Instrumentation (CICT) of the University of Jaén. Another sample, denoted as *Control Coupon*, was stored in a dust-free box to prevent its optical transmittance characteristics from being adversely affected from accidental soiling. Also, this coupon was used as the baseline for each measurement and to check the quality and repeatability of weekly measurements. The soiling transmittance is obtained from the transmittance measurement by means of the following expression:

$$\tau_{soiling}(\lambda) = \frac{\tau_{soiled}(\lambda)}{\tau_{cleaned}(\lambda)} \qquad (11)$$

where $\tau_{soiled}(\lambda)$ and $\tau_{cleaned}(\lambda)$ are, respectively, the spectral transmittance of the *Soiled Coupon* and the *Control Coupon*. As an example, Figure 1 shows the soiling transmittance after one, six and twelve months of outdoor exposure. Independently of the average transmittance, a higher attenuation can be clearly seen at shorter wavelengths for the three cases. Also, it is important to mention that soiling tends to build up in the periods between natural or manual cleaning events, such as rainfalls. For this reason, the amount of soiling and, therefore, the transmittance loss, do not necessarily increase with the number of weeks of outdoor exposure, because soiling can be washed off by rain events. As later discussed in sub-section 4.1, the frequent rainfalls occurred in Month 12 are the reasons why the transmittance in that month is higher than in the drier month 6.

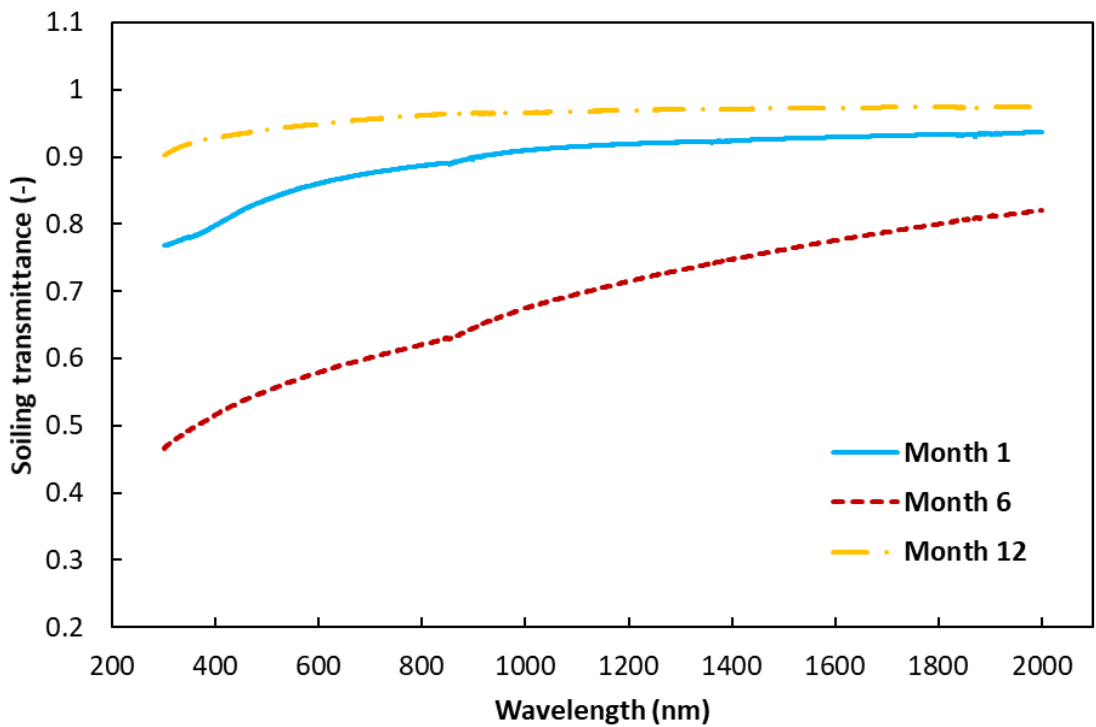

**Figure 1.** Soiling transmittance after one, six and twelve months of outdoor exposure (see sub-section 4.1. for the evolution of soiling across the experimental campaign).

An atmospheric station MTD 3000 from Geonica S.A. located on the rooftop of the CEAEMA recorded the direct normal (DNI), global horizontal (GHI), global normal (GNI) and diffuse horizontal (DHI) irradiances, as well as other relevant parameters such as rainfall, air temperature, wind speed and direction or relative humidity. In addition, a solar spectral irradiance meter (SolarSIM-D2) from Spectrafy Inc. [25] mounted on a high-accurate two axis CPV tracker from

BSQ Solar S.L. was also used to record the spectral distribution of the direct normal irradiance. All these parameters were recorded every 5 minutes and were available over the same period than the transmittance measurements of coupons.

The indexes described in section 2 have been weekly obtained considering the SR or the absorption bands of a typical triple-junction solar cell, the $\tau_{soiling}(\lambda)$ calculated with equation (11) and the spectra recorded during the same day. The basic structure and SR of each subcell of the triple-junction solar cell considered are shown in Figure 2. Table 1 shows the bandwidth interval of each junction, as well as for the whole MJ solar cell considered in this work. The last will also add perspective of the total direct transmittance losses produced by soiling in the short-wavelength spectrum used by solar energy technologies at the present, since the MJ cell virtually covers their whole spectral range. This MJ solar cell has been selected since it represents the most industrialized and used concentrator solar cell nowadays [27]. Moreover, this architecture meets the recommended absorption bands detailed in the IEC 62670-3 standard concerning the spectral evaluation of MJ-based CPV systems, i.e. top = 1.9 eV, middle = 1.4 eV and bottom = 0.7 eV [28]. Hence, the results of the present study can be considered as representative of current concentrator systems.

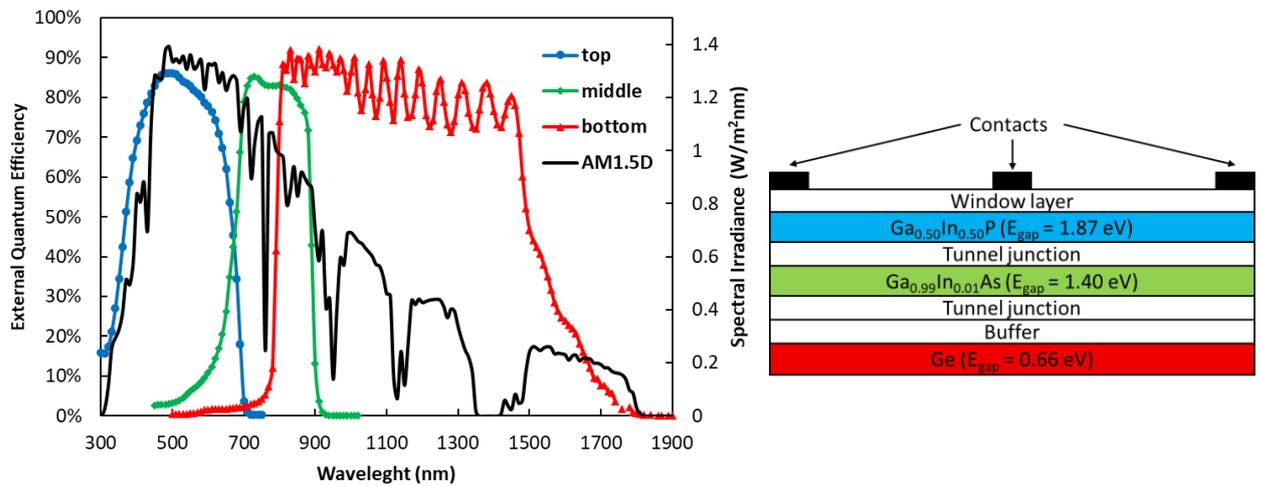

**Figure 2.** External quantum efficiency (left) at 25ºC and schematic (right) of the triple-junction lattice-matched GaInP/GaInAs/Ge reference solar cell considered. The energy gaps ($E_{gap}$) of each subcell are also provided [27].

| Waveband | Material | $\lambda_{min}$ (nm) | $\lambda_{max}$ (nm) |
|---|---|---|---|
| MJ | GaInP/GaInAs/Ge | 300 | 1810 |
| top | GaInP | 300 | 720 |
| middle | GaInAs | 720 | 920 |
| bottom | Ge | 920 | 1810 |

**Table 1.** Wavebands of the lattice-matched GaInP/GaInAs/Ge solar cell considered (see figure 2 for the spectral limits).

Further comments regarding the experimental procedure are worth to mention. The weekly soiling transmittance measurements have been performed three times consecutively. Based on this, the weeks that presented a difference higher than 1% in the $AST_{MJ}$ between the maximum and minimum measurement have been removed from the analysis to avoid noise due to non-uniform soiling or possible measurement errors. Also, in case of cloudy days, the spectra recorded one day before or after the soiling measurements were used to perform the calculations. In this study, days are considered cloudy when the ratio of the total DNI to the total GNI collected during the day is lower than 0.75, i.e. $\Sigma_{day}DNI/\Sigma_{day}GNI < 0.75$, similarly as previously used in [29]. At the end of the experiment, a total of 48 weeks were available to conduct the present investigation.

Finally, this study is based on the direct transmittance of soiling accumulated on the surface of a glass coupon. In case of considering the concentrator optics, other effects such as additional scattering phenomena produced in the rings of Fresnel lenses or on the surface of parabolic mirrors could have an additional impact [30]. Also, the use of secondary optics to improve the acceptance angle (AA) of CPV modules could diminish the losses produced by the scattering [31]. In addition, it is acknowledged that the difference in the angular performance of CPV, which presents AA ≈ 1°, and spectrophotometers, which shows AA ≈ 2°, could impact the results. In this sense, it is important to mention that the use of spectrophotometers is a standard technique to investigate optical properties of CPV systems, including the effects of degradation and soiling [32, 33]. The intention of this work is to contribute to the understanding of the wavelength dependence of the attenuation of soiling and its impact on the spectral performance of MJ-based CPV systems. In this sense, soiling is here considered as a stand-alone optical element, as previously considered by other authors [16]. The investigation of the scattering that may be produced between soiling and the concentrator optics would imply to gather detailed information regarding the size and distribution of soiling across the surface of the CPV system under consideration. In addition, weekly ray-tracing simulations as a function of the recorded input spectra would be necessary. This is out of the scope of this paper since the main intention is to outline general conclusions concerning the spectral impact of soiling valid for any CPV system independently of the concentrator optics used.

## 4. Results

In this section, the main results of the investigation are discussed. Different statistical parameters have been calculated to analyse the relationships among the different variables: the mean absolute percentage error (MAPE), the mean percentage error (MPE) and the determination coefficient ($R^2$). These parameters have been calculated by means of the following expressions:

$$MAPE\ (\%) = \frac{100}{N}\sum_{i=1}^{N}\left|\frac{Z_{modelled}-Z_{measured}}{Z_{measured}}\right| \qquad (12)$$

$$MPE\ (\%) = \frac{100}{N}\sum_{i=1}^{N}\frac{Z_{modelled}-Z_{measured}}{Z_{measured}} \qquad (13)$$

$$R^2 = \left(\frac{\sum_{i=1}^{N}(Z_{measured}-\overline{Z}_{measured})(Z_{modelled}-\overline{Z}_{modelled})}{\sqrt{\sum_{i=1}^{N}(Z_{measured}-\overline{Z}_{measured})^2 \sum_{i=1}^{N}(Z_{modelled}-\overline{Z}_{modelled})^2}}\right)^2 \qquad (14)$$

Where N is the number of samples and Z represents the parameter considered.

4.1. Soiling transmittance

Figure 3 shows the evolution of the $AST_{MJ}$, rainfall, and Particulate Matter 10 (PM10) and 2.5 (PM2.5) over the experimental campaign in order to relate the measured transmittance of soiling with the key weather variables. This could be useful to contribute to future studies regarding this issue, as well as to check the consistency of the recorded data. The $AST_{MJ}$ collected presents an average of 0.891, a maximum of 0.995 and a minimum of 0.585. As shown, the $AST_{MJ}$ is closely related with the rainfall and PM10 trends. In this sense, different events could be highlighted. For instance, the $AST_{MJ}$ strongly decreases on week 5, dropping from 0.897 on week 4 to 0.639, due to an extreme soiling event. On that week, the particulate monitor of the Andalucía air quality monitoring and control network [26], located 1 km from the experimental set-up, recorded an extreme PM10 concentration of 87.5 µg/m$^3$, double of the average concentration during the year. Another noteworthy soiling event occurs from weeks 28 to 36. As can be seen, the $AST_{MJ}$ decreases with a linear behaviour, from a value of 0.977 on week 28 to a minimum value of 0.871 on week 35 ($R^2 = 0.97$). This derate corresponds to the Soiling Rate in the PV field and is produced by the absence of rain over a specific period of time. Finally, another relevant event happens on week 36, where a total of 39.4 mm of rainfall was recorded. As a consequence, the $AST_{MJ}$ drastically increases up to 0.964. This discussion gives evidences regarding the quality of the soiling data collection.

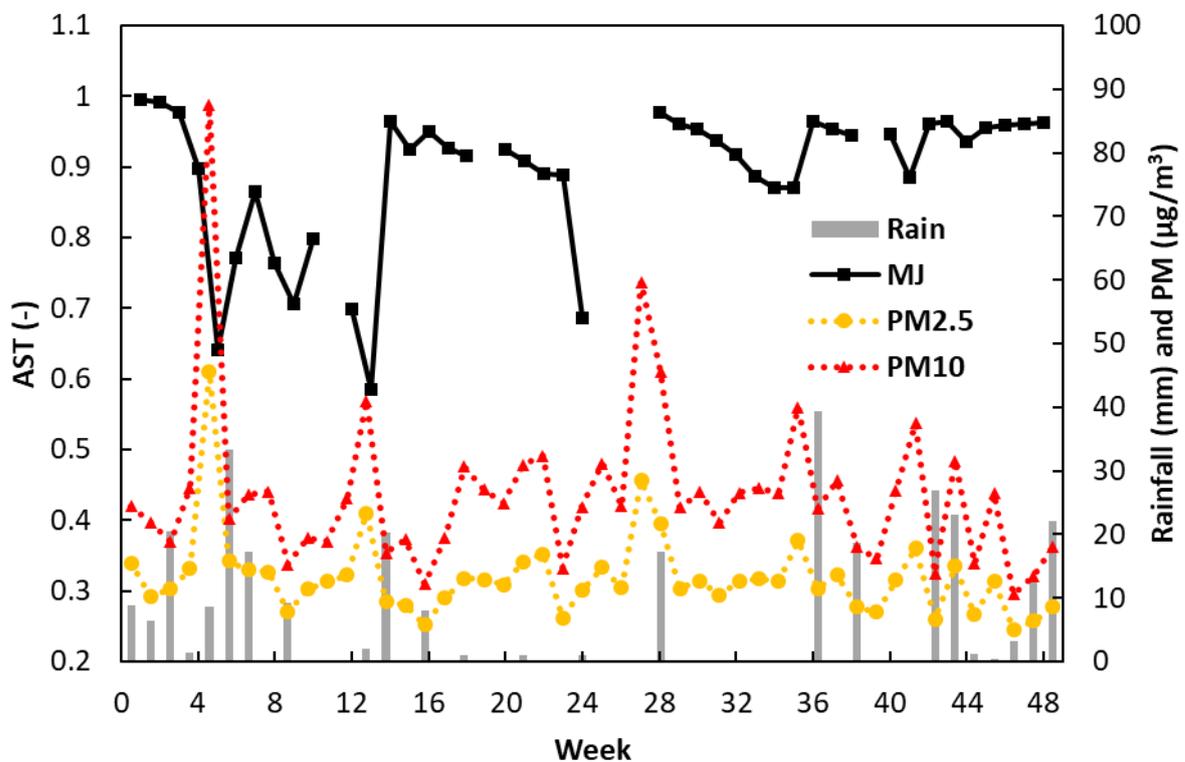

**Figure 3.** Weekly time series of the Average Spectral Transmittance (AST) of soiling across the spectral region of the multi-junction (MJ) solar cell considered, particular matter 2.5 (PM2.5) and 10 (PM10) and rainfall gathered during the one-year outdoor experimental campaign. The daily

values of PM2.5 and PM10 were gathered from a station, located around 1 km from the experimental set-up, of the Andalucía air quality monitoring and control network [26].

Figure 4 shows the AST on each band of the MJ cell. Also, for a better readability and comparison purposes, the $AST_{MJ}$ is again included. As can be seen, all the transmittances qualitatively present the same evolution over time. Despite this, the average transmittance of soiling is systematically different in each spectral band. As shown, $AST_{top}$ presents the lowest transmittance values with an average of 0.829, a maximum of 0.991 and a minimum of 0.480. On the contrary, the $AST_{bot}$ shows the highest transmittance values, with an average of 0.961, a maximum of 1.000 and a minimum of 0.623. Finally, the $AST_{mid}$ presents intermediate values with an average of 0.872, a maximum of 0.992 and a minimum of 0.582. These results are important since they prove that soiling can affect the current balance between the top and middle junctions. The lower impact of soiling on the waveband of the bottom subcell and its typical excess in current are not expected to modify the behaviour of the system. However, the soiling transmittance losses on the top subcell are around 4% higher than in the middle. This is relevant considering that the spectral behaviour of MJ-based CPV systems is mainly driven by the top junction since it is the current-limiting subcell in the majority of the cases [18]. Hence, the inherent spectral nature of soiling and its lower transmittance at shorter wavelengths is expected to introduce additional spectral losses on the systems. This will be further discussed in the next sub-section.

In addition to the discussion above, the relation between the AST of each subcell and in the whole wavelength range of the cell has been investigated. Figure 5 shows the ratios of the average transmittance of the top to the middle subcell ($AST_{top}/AST_{mid}$) and to the bottom subcell ($AST_{top}/AST_{bot}$) plotted against the $AST_{MJ}$. As can be seen, these ratios are not constant and tend to decrease with the broadband reduction of $AST_{MJ}$. This means that the $AST_{top}$ decreases in a larger extent than the AST of the other two junctions as the $AST_{MJ}$ decreases. As can be seen, the $AST_{top}/AST_{mid}$ and $AST_{top}/AST_{bot}$ ratios decrease with the $AST_{MJ}$ following a noticeable linear behaviour. This linear trend indicates that both ratios can be predicted with a low margin of error from the $AST_{MJ}$ by using a simple linear relationship. Indeed, the estimation of the $AST_{top}/AST_{mid}$ ratio shows a MAPE = 1.59% and a MPE = 0.04%, and the estimation of the $AST_{top}/AST_{bot}$ a MAPE = 2.55% and a MPE = 0.28%. The relation of the $AST_{top}/AST_{mid}$ ratio with $AST_{MJ}$ is a remarkable conclusion. This indicates that the more the attenuation caused by soiling in the input irradiance, the more the attenuation in the top subcell respects to the middle. This is relevant since the additional spectral impact caused by soiling is expected to grow with the amount of soiling accumulated on the surface of CPV systems.

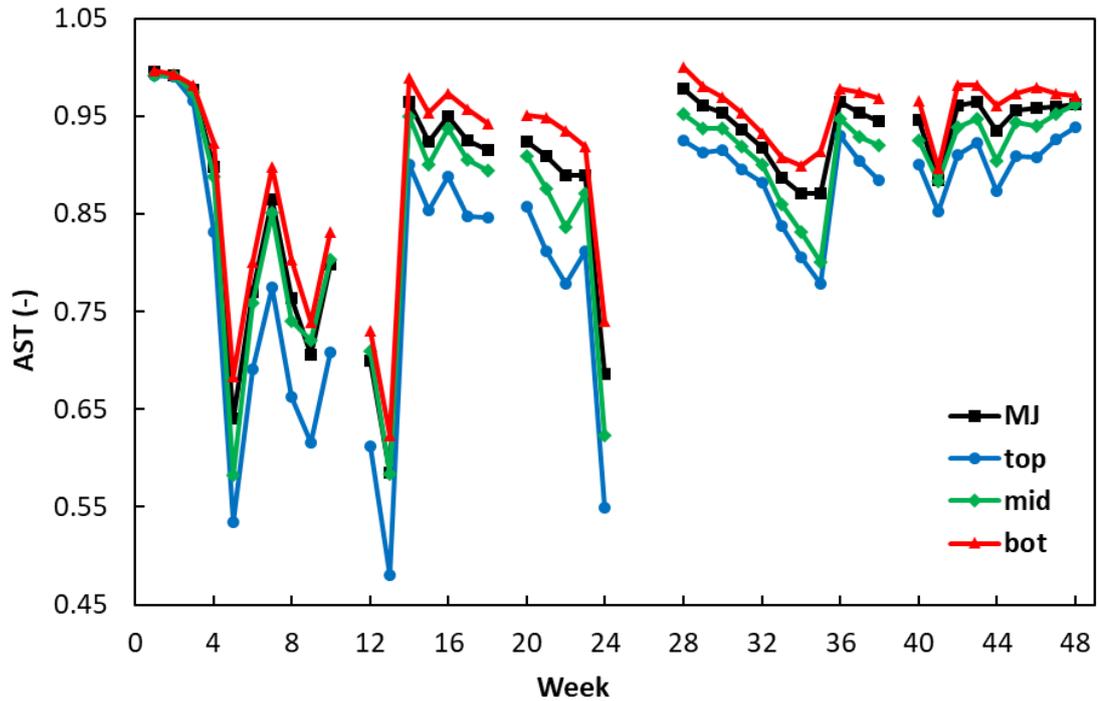

**Figure 4.** Weekly time series of the average spectral transmittance (AST) of soiling across the spectral region of top, middle and bottom junctions, and the multi-junction (MJ) solar cell considered gathered during the one-year outdoor experimental campaign.

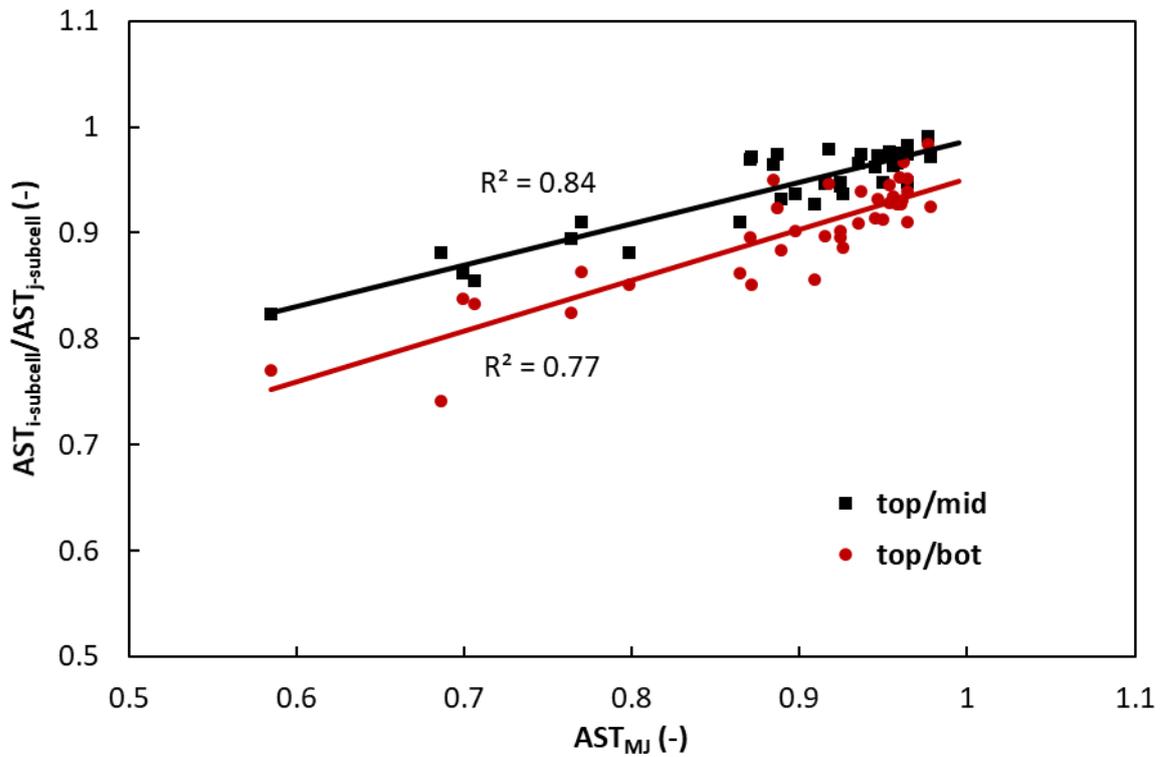

**Figure 5.** Ratio of the AST for the *i*-subcell ($AST_{i\text{-subcell}}$) to the AST for the *j*-subcell ($AST_{j\text{-subcell}}$) versus the AST of soiling across the spectral region of the multi-junction (MJ) solar cell considered.

## 4.2. Soiling impact

Figure 6 shows the evolution of the impact of soiling on the current balance between the top and middle junctions through the SMratio index. In addition, the ratio of the $AST_{top}$ to the $AST_{mid}$ subcells ($AST_{top}/AST_{mid}$) is shown. As can be seen, both magnitudes qualitatively present the same evolution over time. As expected, the presence of soiling tends to decrease the current generated by the top junction to the current generated by the middle junction. The SMratio presents an average value of 0.966, a maximum of 0.999 and a minimum value of 0.878. On the other hand, the $AST_{top}/AST_{mid}$ is more affected by soiling than the SMratio. It presents an average value of 0.947, a maximum value of 0.999 and a minimum value of 0.823. This indicates that the impact of soiling on the current balance among the subcells cannot only be explained considering the attenuation of soiling on each specific spectral band. It is going to be necessary to consider the coupling among the spectral transmittance of soiling, the input spectrum, and the spectral response of each junction. The lower impact of soiling on the SMratio respect to the $AST_{top}/AST_{mid}$ can be understood considering that the irradiance in the waveband of the top subcell is lower than in the middle in the majority of the cases. For instance, it is around 50% lower at AM = 10 [18]. As a consequence, under the standard irradiance, the spectral transmittance of soiling relatively affects the current of the top junction in a lesser extent than the current of the middle junction. Hence, the SMratio is expected to be always higher than the $AST_{top}/AST_{mid}$. Despite this issue, the SMratio can be determined from the $AST_{MJ}$ with a low margin of error. As can be seen in Figure 7, the SMratio shows a linear tendency as the broadband transmittance decreases, an $R^2 = 0.86$, and shows a MAPE = 0.96% and a MPE = 0.02%.

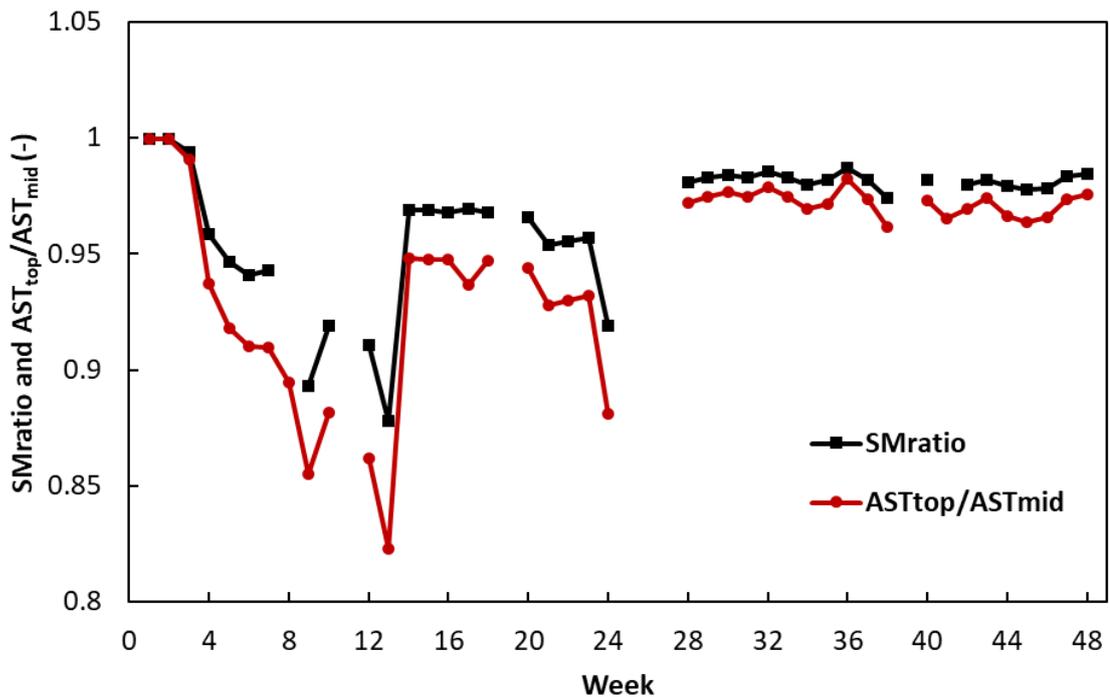

**Figure 6.** Weekly time series of the soiling mismatch ratio (SMratio), and of the ratio of the AST for the top to the AST for the middle subcells ($AST_{top}/AST_{mid}$) gathered during the one-year outdoor experimental campaign.

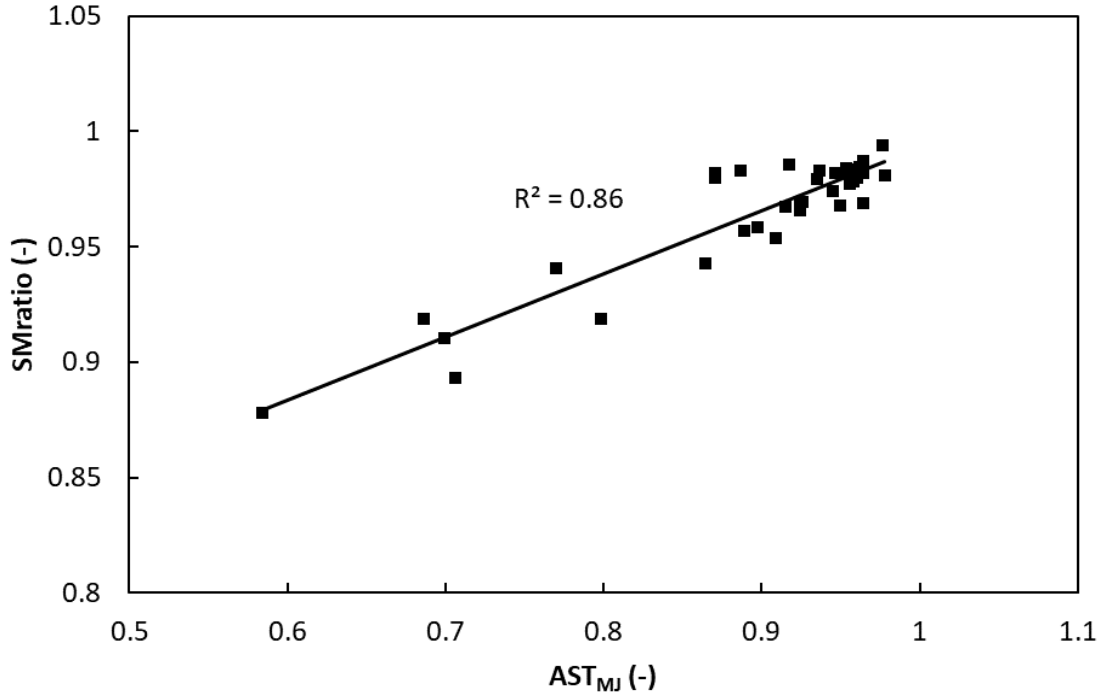

**Figure 7.** Soiling mismatch ratio (SMratio) versus the AST of soiling across the spectral region of the multi-junction (MJ) solar cell considered.

The study of the SMratio above demonstrates that soiling affects the current balance between the top and middle junction of MJ-based CPV systems under real working conditions. In order to investigate how this is translated to soiling losses, Figure 8 shows the average, maximum and minimum values of the SRatio, BSRatio and SSRatio collected during the whole experimental campaign. As can be seen, the SRatio is dominated by the broadband attenuation of soiling. The total soiling losses present an average value of -14.4%, a maximum of -47.5% and a minimum of -0.8%, while the broadband losses show an average of value of -12.7%, a maximum of -44.7% and approximately the same minimum losses. It is important to mention that the annual average BSRatio, i.e. ≈ 0.87, has found to be around 2% lower than the annual average $AST_{MJ}$ discussed in the previous sub-section, i.e. ≈ 0.89. This highlights the importance of considering the actual spectral distribution to remove the broadband losses from the total losses in order to accurately estimate the spectral soiling losses. The additional spectral losses caused by soiling present an average value of -2.0%, a maximum of -7.2% and a minimum of around 0.0%. The contribution to the total losses of the spectral nature of soiling may seem negligible. However, it is worth mentioning that the annual spectral losses of cleaned MJ-based CPV systems are typically around -5%. This is the conclusion of a global investigation conducted at various locations, i.e. Solar Village (Saudi Arabia), Alta Floresta (Brazil), Frenchman Flat (USA), Granada (Spain) and Beijing (China), for different cells, i.e. lattice-matched and metamorphic-mismatched, and optical materials, i.e. poly(methylmethacrylate) and silicon-on-glass(SOG) [15]. This means that soiling can be responsible of an increase of about 40% in the annual spectral losses of CPV. This could affect the cost of electricity and competiveness of MegaWatt size CPV power plants and should be further investigated in future work.

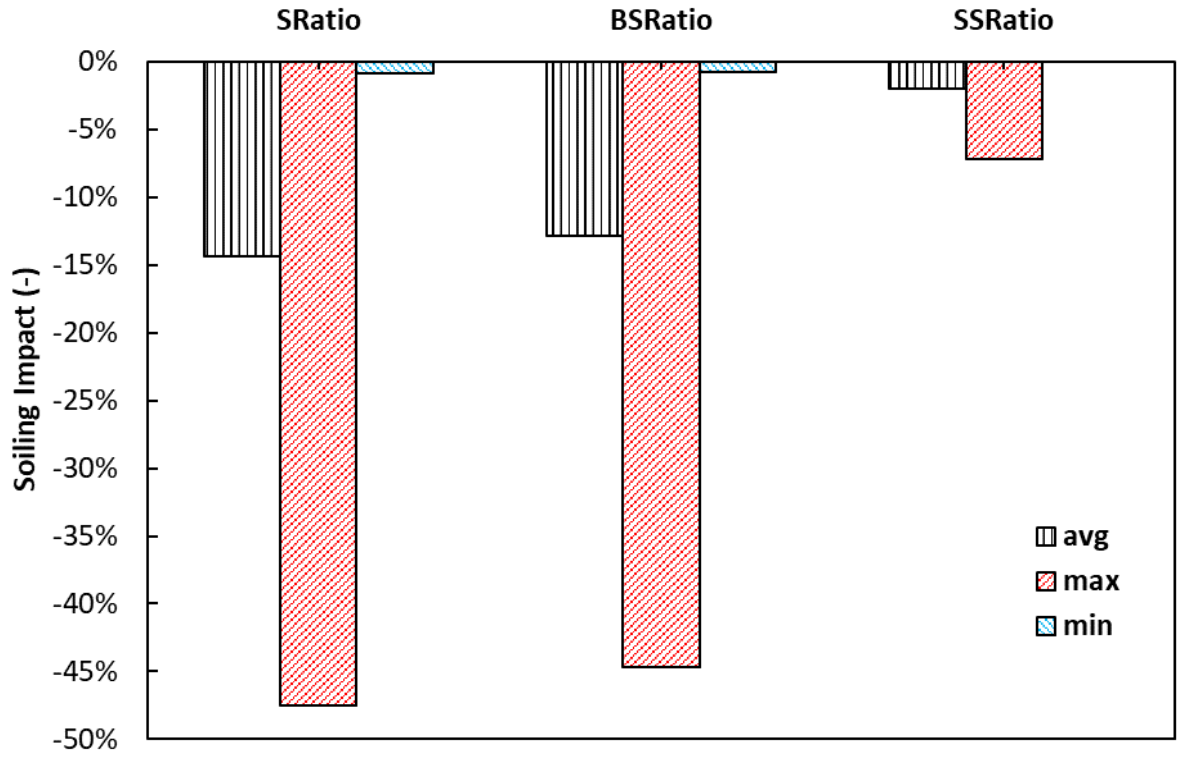

**Figure 8.** Maximum, minimum and average Soiling Ratio (SRatio), Broadband Soiling Ratio (BSRatio) and Spectral Soiling Ratio (SSRatio) obtained during the one-year outdoor experimental campaign.

Figure 9 shows the SRatio, BSRatio and SSRatio versus the AST of soiling on wavelength region of the whole MJ cell. As can be seen, the SRatio and the BSRatio show a clear liner relationship with this magnitude, an $R^2 = 0.98$ and $R^2 = 0.99$, respectively. This indicates that, despite the spectral nature of soiling, the performance of a soiled CPV system is mainly dominated by the broadband attenuation of the irradiance caused by soiling. In this sense, the total soiling losses can be estimated from the $AST_{MJ}$ with a MAPE = 1.48% and a MPE = 0.03%. As expected, the estimation of the broadband losses from $AST_{MJ}$ presents a lower margin of error, i.e. only the coupling between the spectral irradiance and the transmittance of soiling are involved. It shows a MAPE = 1.06% and a MPE = 0.02%. On the contrary, the relation between the SSRatio and the $AST_{MJ}$ shows the poorest linear behavior, i.e. a $R^2 = 0.63$. However, it can also be predicted with a high accuracy with a MAPE = 0.83% and a MPE = 0.01%.

The poorer linear behavior of SSRatio with $AST_{MJ}$ is due to the non-flat spectral transmittance of soiling. However, the value of SSRatio is also going to depend on the distribution of the actual spectral irradiance. In order to illustrate this phenomenon, Figure 10 shows the short-circuit current densities, as well as the DNI, for the top and middle subcells with and without considering soiling for weeks 10 and 35. These weeks have been selected since both present approximately the same soiling spectral losses, i.e. SSRatio = 0.97 (week 10) and 0.96 (week 35), but remarkable $AST_{MJ}$ different values, i.e. $AST_{MJ}$ = 0.79 (week 10) and 0.87 (week 35). Moreover, the $AST_{top}/AST_{mid}$ and SMratio ratios are, respectively, 0.88 and 0.92 in week 10, and 0.97 and 0.98 in week 35. Based on this, it could be expected that week 10 would have higher soiling spectral losses due to the higher reduction of the current on the top junction. The way to elucidate why both weeks present almost the same SSRatio is related to the wavelength distribution of the irradiance. Figure 11 shows the normalized recorded spectral irradiance at noon for the two days used to estimate the different index on those weeks. As can be seen, the spectrum in week 10 is blue-richer than the spectrum in week 35. As a consequence, the current generated by the top junction respect to the middle is higher, i.e. approximately 1% higher than in week 35. As a result of the lower SMratio on week 10, the top and middle junctions stay in current-matching condition at the central hours of the day. In this sense, soiling contributes to improve the spectral performance of the system when the DNI is at maximum. On the contrary, in week 35, soiling reduces the current of both junctions, but it does not contribute to significantly improve the current-matching between the top and middle subcells. This is a relevant conclusion since this phenomenon could help to diminish the soiling losses of MJ-based CPV systems. Indeed, week 10 presents a SRatio approximately 4.5% lower than week 35, i.e. SRatio = 0.755 (week 10) and SRatio = 0.790. However, the $AST_{MJ}$ of week 10 is approximately 8.5% lower than in week 35. This means that week 10 presents approximately 50% lower soiling losses than it would be expected.

Based on the above, it could be concluded the that impact of soiling on the performance of CPV systems could be minimized at locations characterized by a blue-rich spectral distribution. These locations correspond with sites close to the equator, thus with low AM values, and a clear atmosphere, thus low values of aerosols. Hence, locations with high annual irradiation since the AM and aerosols affect the spectral distribution, but also the total irradiance. These sites are, in fact, the ones of interest for CPV technology due their high annual solar resource [34, 35]. This interesting phenomenon could help to diminish the soiling losses and increase the competiveness of CPVs. This is a preliminary conclusion based on the study conducted at a single location. Bearing this in mind, it should be considered and investigated in detail in future works regarding the impact of soiling in the potential of CPV technology. The conclusions drawn from a similar study on the spectral hemispherical transmittance characteristics of soiling in the same site as the present investigation, Jaén, in Spain, were found to be valid also for different locations, such as Golden, in Colorado, and San José, in California. Similarly, the methodology presented in this work should be repeated in different locations to confirm the validity of current findings for various conditions and to generalize these conclusions.

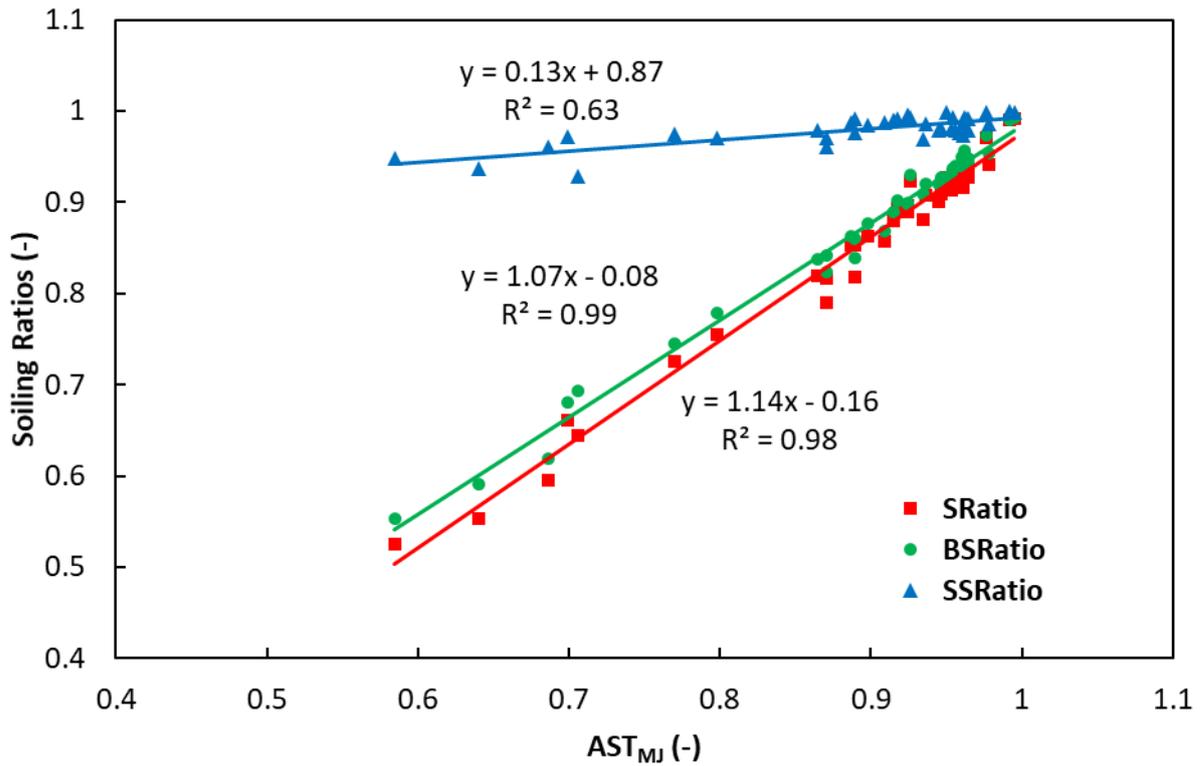

**Figure 9.** Soiling Ratio (SRatio), Broadband Soiling Ratio (BSRatio) and Spectral Soiling Ratio (SSRatio) versus the AST of soiling across the spectral region of the multi-junction (MJ) solar cell considered obtained during the one-year outdoor experimental campaign.

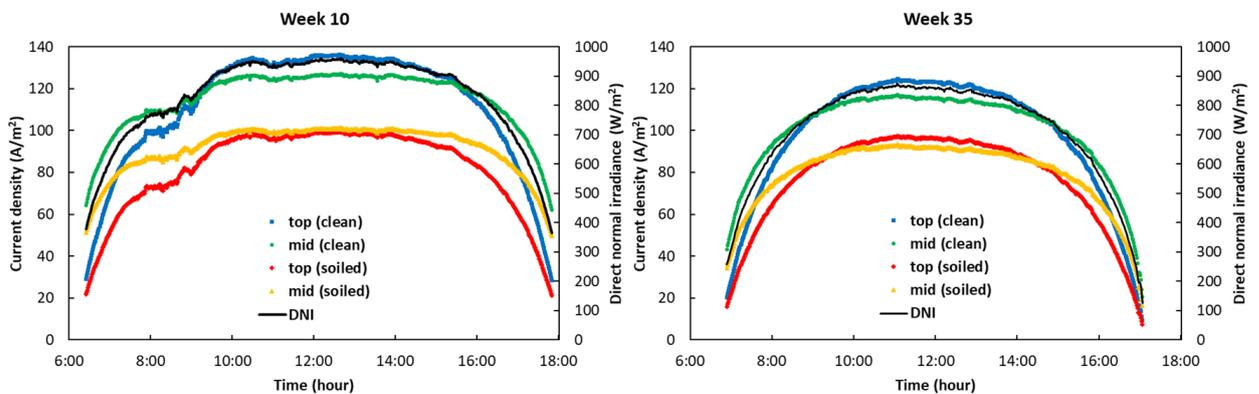

**Figure 10.** Hourly time-series of the short-circuit current densities for the top and middle subcells with and without considering the spectral transmittance of soiling for weeks 10 and 35.

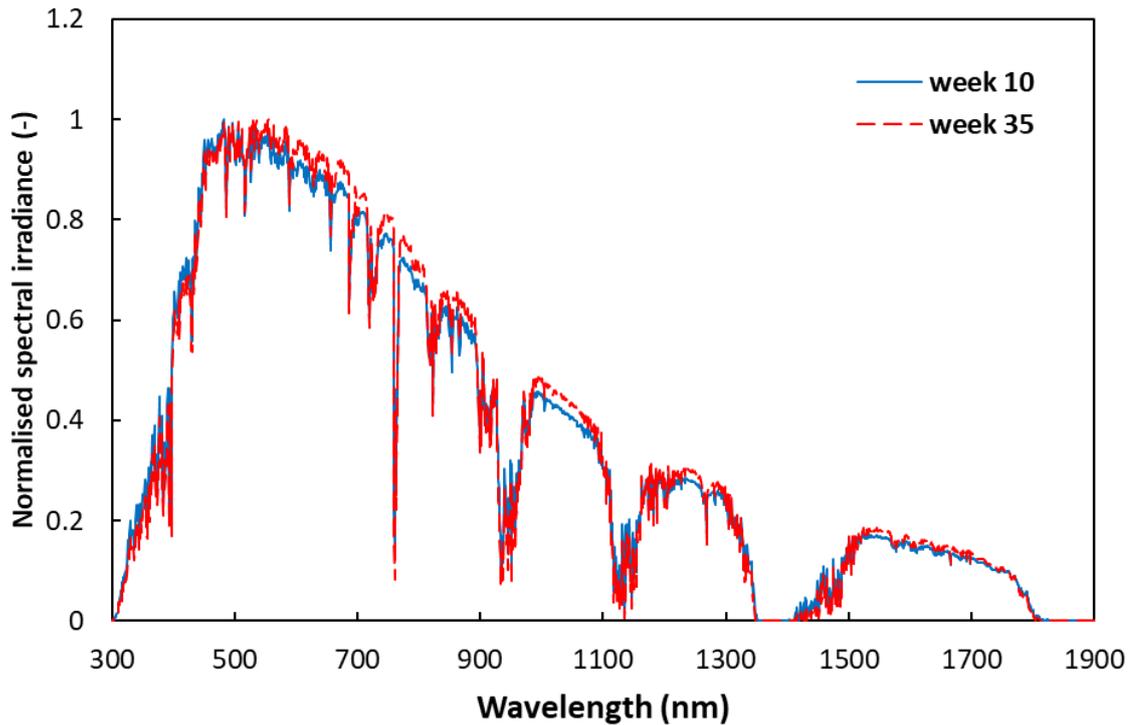

**Figure 11.** Normalised spectral irradiance recorded at noon for weeks 10 and 35.

4.3. Comparison with a four-junction solar cell

This work investigates the spectral effect caused by soiling in CPV systems based on standard triple-junction solar cells. The cell considered is the most widely used in previous work related to the spectral performance of CPV systems. This helps to get general conclusions that could be easily related to other previous studies regarding this issue. For instance, the spectral effects are usually monitored by using component cells made up of the materials considered [12]. However, it is also true that the technology is moving to cells with a higher number of band-gaps to enhance the efficiency and competitiveness of CPV [11]. Bearing this in mind, this sub-section compares the spectral impact of soiling between the typical and a four-junction solar cell. In particular, the record wafer-bonded GaInP/GaAs//GaInAsP/GaInAs cell developed by Fraunhofer ISE has been considered [36].

Figure 12 shows the SR and spectral performance of the four-junction solar cell. Also, the characteristics and results of the triple-junction cell are again included for comparison purposes. As can be seen, both solar cells present approximately the same annual soiling spectral losses. Indeed, the four-junction cell shows a better performance with approximately 0.6% lower average losses. The annual SSRatio is 0.986 for the four-junction solar cell, while it is around 0.980 for the triple-junction cell. This means that, unexpectedly, the shift of the first two subcells in the stack towards shorter wavelengths have shown to reduce the soiling spectral losses. In fact, soiling has found to produce gains up to 2.2%, i.e. a maximum SSRatio of 1.022. This better spectral performance is responsible of a reduction of the total losses soiling of around 1%. The annual SRatio is 0.865, while it is 0.856 for the triple-junction cell. It is also important to mention that the broadband losses are almost equal in both cases since the total absorption band of the two cells is approximately the same. This is a relevant conclusion since it indicates that soiling is expected to have a lower impact on future CPV systems made up of solar cell with a higher number of bandgaps. The diffusion of four-junction cells could help to reduce the soiling losses in CPV systems, and therefore, to increase the competiveness of the technology. MJ cells based on five or more junctions should be also investigated in the future.

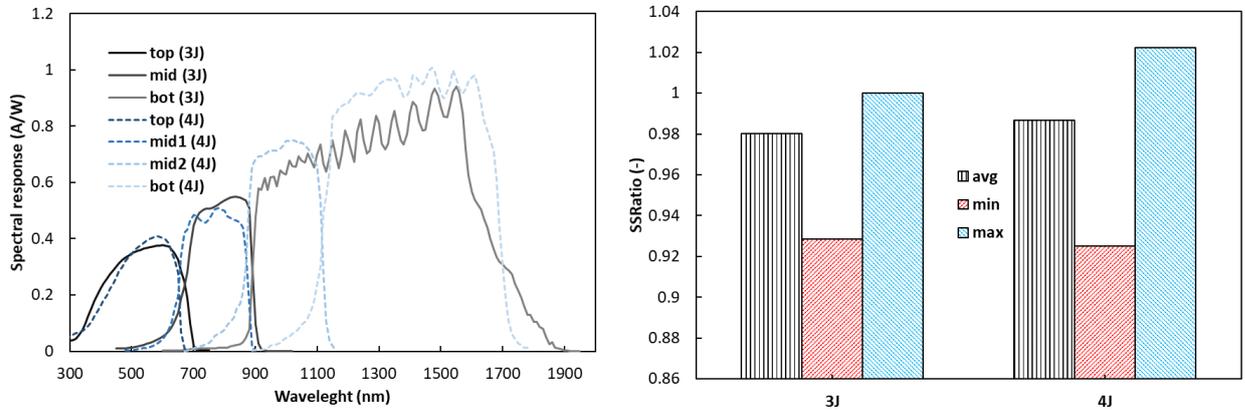

**Figure 12.** Spectral response (left) and Spectral Soiling Ratio (SSRatio) (right) obtained during the one-year outdoor experimental campaign for the lattice-matched GaInP/GaInAs/Ge and the wafer-bonded GaInP/GaAs//GaInAsP/GaInAs solar cells.

## 5. Conclusions

In this paper, the spectral impact of soiling on multi-junction (MJ) cells for CPV applications have been investigated. As a first step, a standard GaInP/GaInAs/Ge cell has been considered. Along with its individual impact on each subcell, soiling also causes changes in the current balance among the subcells, which is also effected by the actual spectral irradiance. Some indexes have been introduced to describe the spectral characteristics of soiling and their impact on a MJ cell performance, and the correlations among them have also been discussed. In general, after one-year experimental investigation conducted in the South of Spain, the attenuation in the transmittance of the top subcell waveband due to soiling has been found to be always greater than the attenuations in the transmittance of the other two subcells wavebands. In addition, the impact of soiling on the transmittance of the top subcell, compared to the transmittance of the other two subcells, has been found to linearly increase with the severity of soiling.

Despite this initial finding, it has been highlighted how the effect of soiling on the current generation of each subcell is also a result of the spectral distribution of the irradiance, which tends to be more intense in the region of the middle subcell. This means that, also if the soiling transmittance attenuation is lower in the mid-subcell waveband than in the top-subcell waveband, its relative impact on the current generated by the mid-subcell can be higher because of the larger amount of incoming light in its waveband.

In this study, soiling has been found to be responsible of annual spectral losses of around 2%, which could be understood as approximately the same average reduction on the power that would deliver a MJ-based CPV system. This magnitude represents about 40% of the total spectral losses, a non-negligible issue for CPV systems. The remaining 60% losses are due to the well-known unavoidable spectral impact of air mass, aerosols and precipitable water. The present work also shows that soiling on CPV could be mitigated in conditions of blue-rich spectral irradiance distribution, thanks to the series-configuration of the subcells in a MJ cell and the higher current generated by the top subcell at these sites during the central hours of day when the irradiance is at maximum. These conditions are common in regions close to the equator and with a clear atmosphere, which have also been previously found to be the most convenient for CPV applications due to their high solar energy resource. These results suggest that the spectral impact of soiling on CPV could be mitigated by the CPV cell structure itself in those locations with favourable conditions for this technology. In addition, it is worth mentioning that a preliminary investigation of a four-junction cell shows that this structure, which have already achieved higher efficiencies than the three-junction cells, can have even a stronger mitigation effect on the spectral losses of soiling, and therefore, in the total soiling impact. A full economic analysis should be conducted, in the future, to understand the actual impact of soiling, and cleanings, on the cost of electricity of CPV, even in comparison with PV.

This work opens the route to understand the spectral impact of soiling on CPV systems. Even if the present work considers data collected over a one-year period in a location that experiences various types of soiling (Saharan dust, olive tree pollen and smoke, and urban particulate matter), future studies should extend this investigation to different locations worldwide. This will help to validate the conclusions and improve our knowledge regarding this issue. Also, the investigation of the spectral impact on real case CPV systems should be addressed. Finally, the relationship between the soiling spectral impact and the crucial parameters that affect the spectrum, i.e. air mass, aerosols and water vapour, should be analysed in future work to improve the current performance modelling techniques.

**Acknowledges**

Eduardo F. Fernández is supported by the "Ministerio de Ciencia, Innovación y Universidades (MICINN)" under the Ramón y Cajal 2017 program. Part of this work was funded through the European Union's Horizon 2020 research and innovation programme under the NoSoilPV project (Marie Skłodowska-Curie grant agreement No. 793120). Part of this work was also funded by MICINN and FEDER funds under the project ENE2016-78251-R.